\documentclass[12pt,twoside,fleqn]{article}
\usepackage{amsmath}
\usepackage{amssymb}
\usepackage{graphicx}
\usepackage{citesort}
\begin{document}

\begin{flushright}
{\small BARI-TH 329/99} \\
{\small DE-FG05-92ER41031-49}
\end{flushright}

\def\obs{{\cal O}}

\thispagestyle{empty}

\begin{center}
{\Large \bf {Further lattice evidence for a large \\[.6cm] 
re-scaling of the Higgs condensate } }

\end{center}
\vspace{1.0cm}
\begin{center}
{\large 
P. Cea$^{1,2}$,
M. Consoli$^{3}$,
L. Cosmai$^{2}$ and 
P. M. Stevenson$^{4}$\\
\vspace{1.0cm}
{\small
$^1$~Dipartimento di Fisica,  Universit\`a di Bari,  via Amendola 173, 
I 70126 Bari, Italy\\[0.05cm] 
$^2$~INFN - Sezione di Bari, via Amendola 173, I 70126 Bari, Italy \\[0.05cm]
$^3$~INFN - Sezione di Catania, Corso Italia 57, I 95129 Catania, Italy 
\\[0.05cm] 
$^4$~Bonner Laboratory, Physics Department, Rice University, Houston TX 77251, 
USA \\[0.05cm]} }

\end{center}
\vspace{0.8cm}
\begin{center}
{\large {\bf Abstract}}
\end{center}
\vspace{0.3cm}
Using a high-statistics lattice simulation of the Ising limit of 
$(\lambda \Phi^4)_4$ theory, we have measured the susceptibility 
and propagator in the broken phase.  We confirm our earlier finding of 
a discrepancy between the field re-scaling implied by the propagator data and 
that implied by the susceptibility.  The discrepancy becomes {\it worse} as 
one goes closer to the continuum limit; thus, it cannot be explained by 
residual perturbative effects.  The data are consistent with an unconventional 
description of symmetry breaking and ``triviality'' in which the re-scaling 
factor for the finite-momentum fluctuations tends to unity, but the 
re-scaling factor for the condensate becomes larger and larger as one 
approaches the continuum limit.  In the Standard Model this changes the 
interpretation of the Fermi-constant scale and its relation to the Higgs mass.

\newpage
\setcounter{page}{1}

\section{Introduction}

   The well-established fact that the continuum limit of $(\lambda \Phi^4)_4$ 
theory is ``trivial'' 
\cite{aizen,froh,sokal,latt,lw,glimm,book}  
is traditionally interpreted using renormalized perturbation theory.
In that picture, the ratio $M_h^2/v_R^2$ of the physical Higgs mass to the 
physical vacuum value $v_R$ is proportional to the renormalized coupling 
$\lambda_R$ and tends to zero as 1/$ \ln$(cutoff).  Since $v_R$ is fixed 
by the Fermi constant $v^2_R \sim 1/(G_F\sqrt{2}) \sim (246 {\rm GeV})^2$, 
the physical Higgs mass is driven to zero in the continuum limit.  To 
avoid that phenomenologically disastrous outcome one must invoke a finite 
cutoff $\Lambda$.  A smaller cutoff value allows a larger Higgs mass, 
but the requirement that $\Lambda \ge 2M_h$ implies an upper bound on $M_h$ 
\cite{lang}.  

   In general, $v_R$ differs from the bare ``Higgs condensate'' value 
measured on the lattice, $v_B \equiv \langle \Phi \rangle$. The two are 
related by a re-scaling factor $Z$: 
\begin{equation}
\label{vphz}
      v_R \equiv \frac{v_B}{\sqrt{Z}}.  
\end{equation}
In the conventional picture $Z = Z_{\rm prop}$ where $Z_{\rm prop}$ is the 
wavefunction-renormalization constant for the propagator of the shifted field 
$\Phi(x)-\langle \Phi \rangle$.  Due to `triviality,' one expects  
$Z_{\rm prop} \to 1$ in the continuum limit, in agreement with the 
perturbative prediction 
$Z_{\mathrm{prop}}=1 + {\cal{O}}(\lambda_{\mathrm{R}})$.  Lattice 
data for the shifted-field propagator confirm that $Z_{\rm prop}$ is 
quite close to 1 \cite{lang}. 

    However, a different interpretation of ``triviality'' has been proposed
\cite{cs,primer,response} in which there are two distinct ``$Z$'s'' in 
the broken-symmetry phase.  The field $\Phi(x)$ must be divided into 
a finite-momentum piece and a zero-momentum (spacetime-constant) piece, 
$\varphi$.  The former re-scales by the usual wavefunction-renormalization 
factor $Z_{\rm prop}$, but the latter re-scales by a different factor, 
$Z_{\varphi}$.  It is $Z = Z_{\varphi}$ that is needed in Eq.~(\ref{vphz}).  
$Z_{\varphi}$ is determined by requiring the physical mass to match the 
second derivative of the effective potential with respect to the renormalized 
$\varphi_R$:
\begin{equation}
M_h^2 = 
\left. \frac{d^2V_{\rm eff}}{d\varphi^2_R} \right|_{\varphi_R=\pm v_R} = 
Z_\varphi 
\left. \frac{d^2V_{\rm eff}}{d\varphi^2_B} \right|_{\varphi_B=\pm v_B} 
= \frac{Z_\varphi}{\chi} , 
\end{equation}
where $\chi$ is the zero-momentum susceptibility.  

     Refs. \cite{cs,primer,response} argue that in a ``trivial'' theory the 
effective potential should be effectively given by the sum of the classical 
potential and the zero-point energy of the shifted fluctuation field, which 
behaves as a free field.  This leads to a $V_{\rm eff}$ that is extremely flat 
in terms of the bare field, implying a logarithmically divergent $Z_{\varphi}$ 
of order $\ln (\Lambda/M_h)$.  Therefore this interpretation of ``triviality'' 
predicts that $Z_{\mathrm{prop}} \to 1$ and $Z_\varphi \to \infty$ in the 
infinite-cutoff limit. 

   A direct test of the ``two $Z$'' picture was reported in our previous 
paper \cite{old}.  There we found a discrepancy between the $Z_{\rm prop}$ 
obtained from the propagator and the $Z_{\varphi}$ obtained from $M_h^2 \chi$. 
Absolutely no sign of such a discrepancy was found in the symmetric phase 
\cite{old}, as expected.  Here we report a substantially refined calculation; 
it involves larger lattices and  a tenfold increase in statistics.  Our 
previous result is confirmed.  

\section{The lattice simulation}

The one-component $(\lambda\Phi^4)_4$ theory   
\begin{equation}
\label{action}
   S =\sum_x \left[ \frac{1}{2}\sum_{\mu}(\Phi(x+\hat e_{\mu}) - 
\Phi(x))^2 + \frac{r_0}{2}\Phi^2(x)  + \frac{\lambda_0}{4} \Phi^4(x)  
\right]    
\end{equation}
becomes in the Ising limit 
\begin{equation}
\label{ising}
   S_{\rm Ising} = -\kappa
\sum_x\sum_{\mu} \left[ 
\phi(x+\hat e_{\mu})\phi(x) +
\phi(x-\hat e_{\mu})\phi(x) \right]    
\end{equation}
with $\Phi(x)=\sqrt{2\kappa}\phi(x)$ and where $\phi(x)$ takes only the 
values $+1$ or $-1$.  

    We performed Monte-Carlo simulations of this Ising action 
using the Swendsen-Wang \cite{SW} cluster algorithm.  
Statistical errors 
can be estimated through a direct evaluation of the integrated autocorrelation 
time~\cite{Madras88}, or by using the ``blocking''~\cite{blocking} or the 
``grouped jackknife''~\cite{jackknife} algorithms.  We have checked that 
applying these three different methods we get consistent results.

    We have measured the following lattice observables: 

(i) the bare magnetization, 
$v_B=\langle |\Phi| \rangle$, 
where $\Phi \equiv \sum_x \Phi(x)/L^4$ is the average field for each 
lattice configuration,

(ii) the zero-momentum susceptibility 
\begin{equation}
\label{suscep}
\chi=L^4 \left[ \left\langle |\Phi|^2 \right\rangle - 
\left\langle |\Phi| \right\rangle^2 \right] ,
\end{equation}

(iii) 
the shifted-field propagator 
\begin{equation}
G(p)= \left\langle \sum_x \exp (ip x) (\Phi(x) - v_B) 
(\Phi(0)- v_B) \right\rangle \, ,
\end{equation}
where $p_{\mu}=\frac{2\pi}{L}n_{\mu}$ 
with $n_{\mu}$ being a vector with integer-valued components, not all zero.  
We shall compare the lattice data for $G(p)$ to the 2-parameter formula 
\begin{equation}
\label{gpform}
G_{\rm fit}(p)= \frac{Z_{\rm prop}}{\hat{p}^2 + m^2_{\rm latt}} ,
\end{equation}
where $m_{\rm latt}$ is the mass in lattice units and 
$\hat{p}_{\mu}=2 \sin \frac{p_{\mu}}{2}$.  If ``triviality'' is true, then 
this form should give a better and better description of the lattice data 
as we approach the continuum limit; also, $Z_{\rm prop}$ should tend to 
unity. 

    Another way to determine the mass is to use the method of ``time-slice'' 
variables described in ref.\cite{montweisz} (see also \cite{montmunster} 
pp. 56).  To this end let us consider a lattice with 3-dimension $L^3$ and 
temporal dimension $L_t$ and the two-point correlator 
\begin{equation}
\label{corr}
C_1(t,0; {\bf k})\equiv \langle 
S_c(t;{\bf k})S_c(0;{\bf k})+
S_s(t;{\bf k})S_s(0;{\bf k}) \rangle _{\rm conn} ,
\end{equation}
where
\begin{equation}
\label{cos}
S_c(t; {\bf k})\equiv \frac{1}{L^3} \sum _{ { \bf x} } \phi({\bf x}, t)
\cos ({\bf k} \cdot {\bf x}) ,
\end{equation}
\begin{equation}
\label{sin}
S_s(t;{ \bf k})\equiv \frac{1}{L^3} \sum _ {{\bf x}} \phi({\bf x}, t)
\sin ({\bf k} \cdot {\bf x}) .
\end{equation}
Here, $t$ is the Euclidean time; ${\bf x}$ is the spatial part of the site 
4-vector $x^{\mu}$; ${\bf k}$ is the lattice momentum 
${\bf k}=(2\pi/L) (n_x,n_y,n_z$), with $(n_x,n_y,n_z)$ non-negative integers; 
and $\langle ...\rangle_{\rm conn}$ denotes the connected expectation 
value with respect to the lattice action, Eq.~(\ref{ising}). In this way, 
parameterizing the correlator $C_1$ in terms of the energy $\omega_k$ as 
\begin{equation}
\label{fitcor}
C_1(t,0;{\bf k})= A \, [ \, \exp(-\omega_k t)+\exp(-\omega_k(L_t-t)) \, ] \,,
\end{equation}
the mass can be determined through the lattice dispersion relation
\begin{equation}
\label{disp}
m^2_{\rm TS}({\bf k}) = ~2 (\cosh \omega_k  -1)~~ -~~2 \sum ^{3} _{\mu=1}~ 
(1-\cos k_\mu) \,.
\end{equation}
In a free-field theory $m_{\rm TS}$ is independent of {\bf k} and 
coincides with $m_{\mathrm{latt}}$ from Eq.~(\ref{gpform}).

\section{Numerical results: symmetric phase}

As a check of our simulations we started our analysis at  $\kappa=0.0740$
in the symmetric phase, where high-statistics results by Montvay and 
Weisz~\cite{montweisz} are available.   
In Fig.~1 we report the data for the scalar propagator suitably re-scaled
in order to show the very good quality of the fit to Eq.~(\ref{gpform}). 
The 2-parameter fit gives $m_{\mathrm{latt}}=0.2141(28)$ and 
$Z_{\rm prop} = 0.9682(23)$.
The value at zero-momentum is defined as  
$Z_\varphi \equiv m^2_{\rm latt} \chi = 0.9702(91)$.
Notice the perfect agreement between $Z_\varphi$ and  $Z_{\rm prop}$.

In Fig.~2 we show the values of the time-slice mass Eq.~(\ref{disp}) at 
several values of the  3-momentum and the corresponding result of 
Ref.~\cite{montweisz}.  The shaded area corresponds to the value 
$ m_{\rm latt} = 0.2141(28)$ obtained from the fit to the propagator data. 
We see that $m_{\rm TS}$ is indeed independent of {\bf k} and agrees 
well with $m_{\rm latt}$.  

Thus, our analysis of the symmetric phase is in good agreement
with Ref.~\cite{montweisz} and shows the expected ``trivial'' behaviour.  
Note that our result for $Z_{\rm prop} \simeq Z_\varphi$ is in excellent 
agreement with the 1-loop renormalization group prediction 
$ Z_{\rm pert} = 0.97(1) $~\cite{lw}. 

\section{Numerical results: broken phase}

We now choose for $\kappa$ three successive values, 
$\kappa = 0.076, 0.07512, 0.07504$, lying just above the critical 
$\kappa_c \simeq 0.0748$ \cite{montweisz}.  Thus, we are in the broken phase 
and approaching the continuum limit where the correlation length $\xi$ becomes 
much larger than the lattice spacing.

We used lattice sizes $L^4$ with
$L=20, 32, 32$, respectively. 
This should ensure that finite-size effects are sufficiently under control,
since in all cases $L/\xi > 5$ \cite{L/4,montweisz}.  
Finite-volume tunneling effects 
\cite{jansen,jansen1} should also be negligible, as we explain in the 
appendix. 
Finally, we also repeated the measurements for $\kappa=0.076$ on an
$L=32$ lattice (with less statistics) to confirm directly the absence of
finite-size effects in this case.  
After discarding 10K sweeps for thermalization, we have performed 
500K sweeps (ten times more than in our earlier calculation \cite{old}); 
the observables were measured every 5 sweeps. 
Our results for the magnetization and the susceptibility are reported in 
Table 1.  We note that at $\kappa = 0.076$ our values are in excellent 
agreement with the corresponding results of Jansen {\it et al} \cite{jansen}.  

    The  data for the re-scaled propagator are reported in Figs. 3---5.  
Unlike Fig.~1, the fit to Eq.~(\ref{gpform}), though excellent at higher 
momenta, does not reproduce the lattice data down to zero-momentum. 
Therefore, in the broken phase, a meaningful determination of $Z_{\rm prop}$ 
and $m_{\rm latt}$ requires excluding the lowest-momentum points from the 
fit.  Fig. 6 shows how the chi-squared per d.o.f. of the fit improves, and 
the fitted value of $m_{\rm latt}$ stabilizes, as low-momentum points are 
excluded from the fit.  Our numbers for $m_{\rm latt}$ and $Z_{\rm prop}$ 
are reported in Table 2.  $Z_{\rm prop}$ is indicated in Figs.~3---5 by 
a dashed line.  

    The fitted $Z_{\rm prop}$ is slightly less than one.  This fact is 
attributable to residual interactions since we are not exactly at the 
continuum limit, so that the theory is not yet completely ``trivial.''  
This explanation is reasonable since we see a tendency for $Z_{\rm prop}$ 
to approach unity as we get closer to the continuum limit.  Moreover, 
we find good agreement between our result, $Z_{\mathrm{prop}}=0.9321(44)$, 
and the L\"uscher-Weisz perturbative prediction 
$Z_{\rm pert}=0.929(14)$~\cite{lw} at $\kappa = 0.0760$.  The comparison 
$Z_{\mathrm{prop}}=0.9566(13)$ with $Z_{\mathrm{pert}}=0.940(12)$ at 
$\kappa=0.07504$ is also fairly good.  

   The quantity $Z_{\varphi}$ is obtained from the product 
$m_{\rm latt}^2 \chi$ and is shown in Figs. 3---5, as a point at $\hat{p}=0$.  
According to conventional ideas $Z_{\varphi}$ should be {\it the same} as the 
wavefunction-renormalization constant, $Z_{\rm prop}$, but clearly it is 
significantly larger.  Note that there was no such discrepancy in Fig. 1 
for the symmetric phase.  

     Our data show that the discrepancy gets {\it worse} as we approach 
the critical $\kappa$.  Fig. 7 shows that $Z_{\varphi}$ grows rapidly 
as one approaches the continuum limit (where $m_{\rm latt} \to 0$).  
Thus, the effect cannot be explained by residual perturbative 
${\cal{O}}(\lambda_{\mathrm{R}}) $ effects that might cause $G(p)$ to 
deviate from the form in (\ref{gpform}); such effects die 
out in the continuum limit, according to ``triviality.''   

     The results accord well with the ``two $Z$'' picture in which, as 
we approach the continuum limit, we expect to see the zero-momentum point, 
$Z_\varphi \equiv m_{\rm latt}^2 \chi$, become higher and higher.  In the 
continuum limit the effect should be confined to the zero-momentum point, 
but it seems that, away from the continuum limit, the effect ``spills over'' 
into the very low momentum modes.  This ``spillover'' would explain 
why the propagator deviates from free-field form at low $\hat{p}$, 
necessitating the exclusion of the lowest few $\hat{p}$ points to get a good 
fit to (\ref{gpform}).  As we approach the continuum theory  the 
deviation from the $Z_{\rm prop}/(\hat{p}^2 + m^2_{\rm latt})$ form should 
become concentrated in a smaller and smaller range of $\hat{p}$, but with 
a larger and larger spike at $\hat{p}=0$.  In this limit the shape 
should become a discontinuous function that is infinite at $\hat{p}=0$ 
and equal to 1 for all non-zero $\hat{p}$.  The sequence of pictures in 
Figs. 3---5 is quite consistent with this expectation.  

    The time-slice mass $ m_{\rm TS}({\bf k})$ also shows distinctive 
behaviour at low momentum, as seen in Fig. 8 for $\kappa=0.076$.  
Except at low momentum the time-slice mass agrees well with $m_{\rm latt}$.  
However, the time-slice mass at zero momentum, $ m_{\rm TS}(0)$, is very 
significantly smaller than $m_{\rm latt}$ (see Table 3).  At $\kappa=0.076$ 
our result for $m_{\rm TS}(0)$ is in very good agreement with the 
corresponding result in Ref.~\cite{jansen}.  The ratio 
$\frac{ m_{\rm TS}(0)}{m_{\rm latt}}$ is shown in Fig. 9 as a function of 
$m_{\rm latt}$.  The trend is inverse to that in Fig. 7 for $Z_{\varphi}$.  

      Our interpretation of the data regards $m_{\rm latt}$ as the true 
particle mass and views $m_{\rm TS}(0)$ as a symptom of the distinct 
dynamics of the zero-momentum mode.  An alternative interpretation 
might be to regard $m_{\rm TS}(0)$ as the true mass.  However, we see 
two serious objections to that interpretation.  Firstly, it would entail 
another ``$Z$'' factor, $Z_0 \equiv \chi m_{\rm TS}(0)^2$, which turns out 
to be roughly $0.88$ in all three cases.  This $Z_0$, unlike our 
$Z_{\rm prop}$, would not agree well with the L\"uscher and Weisz predictions.  
(In fact such a discrepancy in the $\kappa=0.076$ case was noted by Jansen 
{\it et al} \cite{jansen}.)  Secondly, if  $m_{\rm TS}({\bf k})$ depends 
sensitively on ${\bf k}$, as it does near ${\bf k} =0$, then the dispersion 
relation is not well approximated by the (lattice version of the) usual form 
$\omega_k = \sqrt{{\bf k}^2 + {\rm const.}}$; in that case the very notion 
of ``mass'' becomes problematic.  At larger ${\bf k}$, where 
$m_{\rm TS}({\bf k})$ does become insensitive to ${\bf k}$, it agrees 
well with $m_{\rm latt}$.  

\section{Conclusions} 

    Our data clearly reveal unconventional behaviour at and near zero 
momentum in the broken phase.  Well away from zero momentum the propagator 
data fit the ``trivial'' form $Z_{\rm prop}/(\hat{p}^2+m^2_{\rm latt})$, 
with $Z_{\rm prop}$ in accord with perturbative predictions \cite{lw}. 
However, the data deviate from this form at very low momentum, 
presaging an even more dramatic difference between $Z_{\rm prop}$ 
and $Z_{\varphi} \equiv m^2_{\rm latt} \chi$, where $\chi$ is the 
zero-momentum susceptibility.  This difference gets rapidly {\it larger} 
as one gets closer to the continuum limit, $m_{\rm latt} \to 0$; see Fig. 7. 
Because of this fact it is not possible to explain the effect in terms of 
residual perturbative interactions (even if one were to use a different 
value of the mass); perturbative effects should die out as one approaches 
the continuum limit, according to ``triviality.''  Finite-volume effects, 
including tunneling, should be negligible (see appendix).  We note that at 
$\kappa=0.076$, $L=20$ our measurements are in complete agreement with 
Jansen {\it et al} \cite{jansen} wherever a direct comparison is possible.  

    Our results are consistent with the unconventional picture of 
``triviality'' and spontaneous symmetry breaking of 
refs.~\cite{cs,primer,response} in which $Z_{\varphi}$ diverges 
logarithmically, while $Z_{\rm prop} \to 1$ in the continuum limit.  
In this picture the Higgs mass $M_h$ can remain finite in units of the 
Fermi-constant scale $v_R$, even though the ratio $M_h/v_B$ goes to zero.  
The Higgs mass is then a genuine collective effect and $M_h^2$ is {\it not} 
proportional to the renormalized self-interaction strength.  If so, then 
the whole subject of Higgs mass limits is affected.  In view of the 
importance of the issue, both for theory and phenomenology, we hope and 
expect that our lattice results will be checked (and/or challenged) by other 
groups.

\vskip 20pt
{\bf Acknowledgements}
     We thank K. Jansen and J. Jers\'ak for useful discussions and 
correspondence concerning the results of Refs. \cite{jansen,jansen1}.  

     This work was supported in part by the U.S. Department of Energy under 
Grant No. DE-FG05-92ER41031 and in part by the Istituto Nazionale di Fisica 
Nucleare. 

\vskip 40pt

\section*{Appendix:  Tunneling effects}

In this appendix we argue that tunneling effects are negligible in our data.  
Finite-volume tunneling effects in the Ising model have been studied very 
thoroughly by Jansen {\it et al} \cite{jansen,jansen1}.  In finite volume 
there is mixing between the $+v$ and $-v$ `ground states' and the true 
ground state is the symmetric combination with energy $E_{0s} \equiv 0$.  
The antisymmetric combination is slightly higher in energy by an amount 
$E_{0a}$ which depends on the lattice size $L$:
\begin{equation}
          E_{0a} \approx C L^{1/2} \exp{(-\sigma L^3)}.
\end{equation}
This formula comes from a semiclassical instanton-type calculation, 
which also yields theoretical formulas for $C$ and $\sigma$ 
(Ref. \cite{jansen}, Sect. 4.3).  For the case $\kappa = 0.076$, 
Jansen {\it et al} also measured $E_{0a}$ on small lattices ($L \le 10$) 
\cite{jansen1} and obtained values $\sigma = 0.00358(2), C=0.101$ that agree 
well with their theoretical formulas.  Using their results we can estimate 
$E_{0a}$ in our three cases ($\kappa=0.076, 0.07512, 0.07504$ with 
$L=20,32,32$, respectively).  We find $E_{0a} = 2 \times 10^{-13}, 
3 \times 10^{-8}$, and $10^{-5}$, respectively.  

    For the first excited state (in the zero 3-momentum sector) there are 
also two nearly degenerate states, $E_{1s}$ and $E_{1a}$.  The difference 
$E_{1a}-E_{1s}$ is considerably larger than $E_{0a}$.  Ref. \cite{jansen1}'s 
data for $\kappa=0.076$ shows that the ratio of $E_{1a}-E_{1s}$ to $E_{0a}$ 
is around 6 for $L=8$ and around 4 for $L=10$.  It therefore seems reasonable 
to assume that, for larger lattices, $E_{1a}-E_{1s}$ is not more than 4 times 
$E_{0a}$.  (For $\kappa=0.076, L=20$ this implies 
$E_{1a}-E_{1s} \sim 10^{-12}$, so we are fully justified in averaging 
Ref. \cite{jansen}'s results $E_{1s}=0.3914(12)$ and $E_{1a} = 0.3909(14)$ 
in order to compare with our $m_{TS}(0)$ result in Table 3.)

    Tunneling effects lead to a modification of the formulas for $\chi$ 
and the two-point correlator $C_1$.  However, the effect of these 
modifications is tiny in our cases.  When expanded for small $E_{0a}$, the 
non-trivial factor involved in $\chi$ (see Ref. \cite{jansen}, Eq. (24)) 
is seen to differ from unity only by an amount of order $E_{0a}^2 T^2$, where 
$T$ is the lattice's temporal size.  (There is a not-immediately-obvious 
cancellation of the terms linear in $E_{0a}$.)  This difference from unity 
is tiny ($10^{-23}$, $10^{-12}$, $10^{-7}$, in our three cases). Therefore 
we believe that tunneling effects are too small to show up in our data.

\newpage

\begin{table}
\renewcommand{\arraystretch}{2}
\begin{center}
\begin{tabular}{ccclll}
\hline \\[-0.95cm]
\hline
Ref.               &$L_{\rm{size}}$  &$\#$ sweeps     &$\kappa$  
&$\frac{v_B}{\sqrt{2\kappa}}$ &$\frac{\chi}{2\kappa}$      \\ \hline
Our data             &$20^4$            &$5\times10^5$   &0.076 
    &0.3015(1)         &37.71(22)          \\
Ref.~\cite{jansen}  &$20^4$  &$7.5\times10^6$ &0.076   
  &0.30158(2)         &37.85(6)                \\
Our data             &$32^4$            &$2.5\times10^5$   &0.076    
 &0.3015(1)         &37.70(31)                 \\
Our data             &$32^4$            &$4\times10^5$   &0.07512    
 &0.1617(1)         &193.1$\pm$1.7                 \\
Our data             &$32^4$            &$5\times10^5$   &0.07504   
  &0.13822(12)         &293.38$\pm$2.86             \\
\hline \\[-0.95cm]
\hline
\end{tabular}
\caption{ The lattice data for the magnetization and the 
susceptibility.}
\end{center}
\label{table:I}
\end{table}

\newpage
\begin{table}
\renewcommand{\arraystretch}{2}
\begin{center}
\begin{tabular}{ccllll}
\hline \\[-0.95cm]
\hline
              $L_{\rm{size}}$  &$\#$ sweeps     &$\kappa$  
&$m_{\rm{latt}}$ & $Z_{\rm prop}$   &$Z_{\varphi}$ \\ \hline
$20^4$            &$5\times10^5$   &0.076 
    &0.42865(456)         &0.9321(44) &1.0531(232)          \\
$32^4$            &$2.5\times10^5$   &0.076    
 &0.42836(500)     &0.9312(27)  &1.0516(260)          \\
$32^4$            &$4\times10^5$   &0.07512    
 &0.20623(409)     &0.9551(21)  &1.2340(502)          \\
$32^4$            &$5\times10^5$   &0.07504   
  &0.17229(336)        &0.9566(13)     &1.307(52)     \\
\hline \\[-0.95cm]
\hline
\end{tabular}
\caption{Our values for $m_{\rm latt}$ and $Z_{\rm prop}$ as obtained 
from a fit to the lattice data for the propagator (see Figs.3---5). We also 
report $Z_{\varphi}\equiv m^2_{\rm latt} \chi$ where the
$\chi$'s are given in Table 1.}
\end{center}
\label{table:II}
\end{table}

\begin{table}
\renewcommand{\arraystretch}{2}
\begin{center}
\begin{tabular}{ccclll}
\hline \\[-0.95cm]
\hline
Ref.               &$L_{\rm{size}}$  &$\#$ sweeps     &$\kappa$  
&$ m_{\rm TS}( 0 )$ &$ m_{\rm latt}$      \\ \hline
Our data             &$20^4$            &$5\times10^5$   &0.076 
    &0.388(10)         &0.42865(456)          \\
Ref.~\cite{jansen}  &$20^4$  &$7.5\times10^6$ &0.076   
  &0.3912(12)         & --                \\
Our data             &$32^4$            &$4\times10^5$   &0.07512    
 &0.1737(24)         &0.20623(409)                 \\
Our data             &$32^4$            &$5\times10^5$   &0.07504   
  &0.1419(17)         &0.17229(336)            \\
\hline \\[-0.95cm]
\hline
\end{tabular}
\caption{ The lattice data for the zero-momentum time-slice mass compared 
with the mass from the finite-momentum propagator.  The result of 
Ref.~\cite{jansen} is an average of the two values $E_{1s}$ and $E_{1a}$ 
reported in their Table 4. }
\end{center}
\label{table:III}
\end{table}

\clearpage

\section*{FIGURE CAPTIONS}

\renewcommand{\labelenumi}{Figure \arabic{enumi}.}
\begin{enumerate}
\item The lattice data for the re-scaled propagator at $\kappa=0.0740$ in the
symmetric phase. The zero-momentum full point is defined as 
$Z_\varphi=m^2_{\mathrm{latt}}\chi$.
The dashed line indicates the value of $Z_{\mathrm{prop}}$.
\item The data for the time-slice mass Eq.(12) at different values of the
3-momentum. The shaded area represents the value $ m_{\rm latt} = 0.2141(28)$
obtained from the fit of the propagator data. The black square is the result
of Ref.~\cite{montweisz}.
\item The lattice data for the re-scaled propagator at $\kappa=0.076$.
The zero-momentum full point is defined as $Z_\varphi=m^2_{\mathrm{latt}}\chi$.
The very low momentum region is shown in the inset. 
The dashed line indicates the value of $Z_{\mathrm{prop}}$.
\item The same as in Fig. 3 at $\kappa=0.07512$.
\item The same as in Fig. 3 at $\kappa=0.07504$.
\item The values of the reduced chi-square and of the fitted $ m_{\rm latt}$ 
as a function of the number of low-momentum points excluded from the fit to 
Eq.(\ref{gpform}).
\item The two re-scaling factors $Z_{\mathrm{prop}}$ and
$Z_\varphi$ as a function of $m_{\mathrm{latt}}$.  (The continuum limit 
corresponds to $m_{\rm latt} \to 0$.)
\item The time-slice mass $m_{\rm TS}({\bf k})$ for several values of the 
spatial momentum.  The black square at zero momentum is the result of Ref. 
\cite{jansen}.  The shaded area represents the value 
$m_{\rm latt}=0.42865(456)$ obtained from the fit to the propagator data.  
\item The ratio $\frac{m_{\rm TS}(0)}{m_{\rm latt}}$ 
as a function of $m_{\rm latt}$.  
\end{enumerate}

\clearpage
\begin{figure}[t]
\begin{center}
FIGURE 1
\end{center}
\label{Fig1}
\begin{center}
\includegraphics[clip,width=0.9\textwidth]{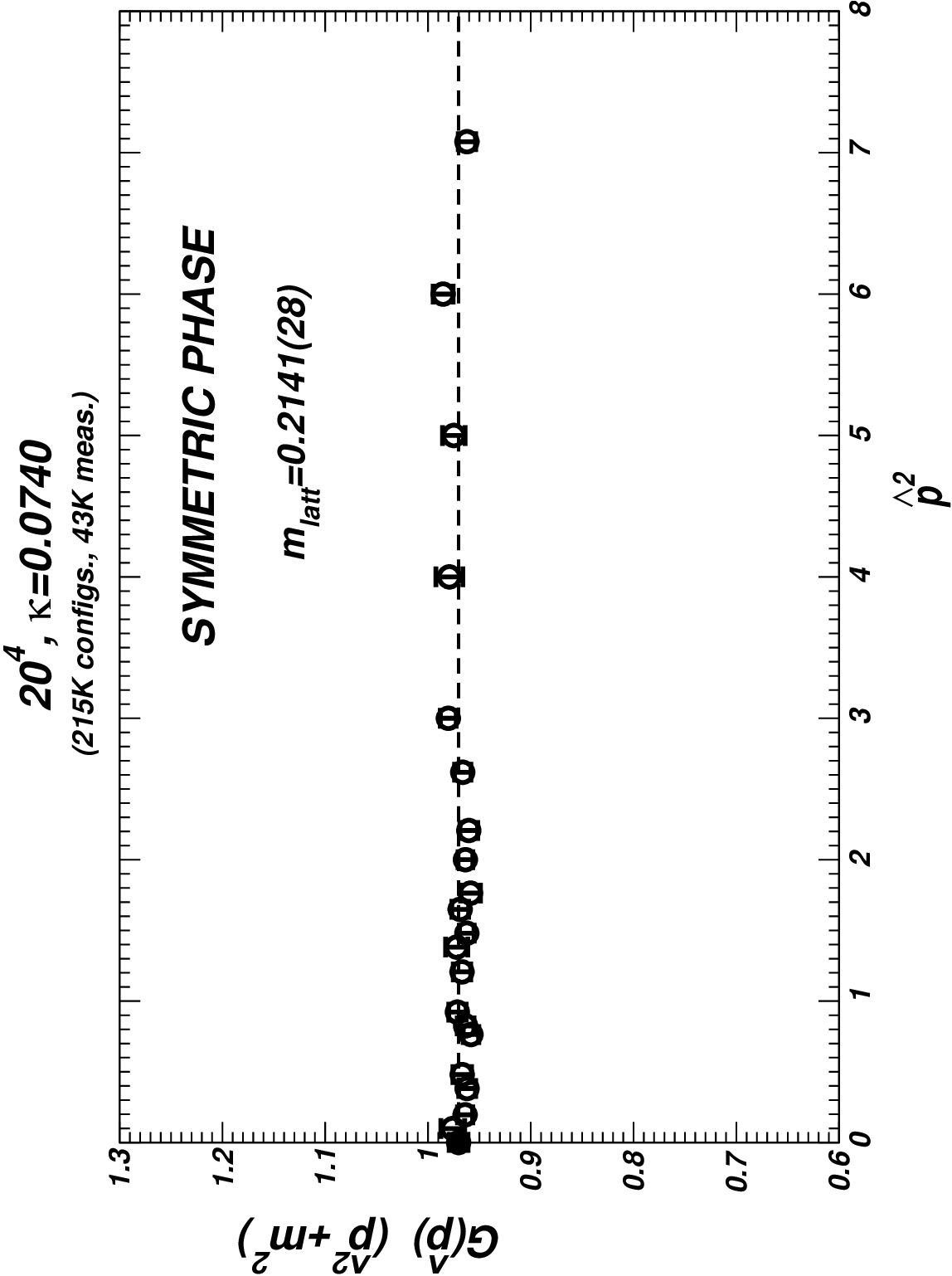}
\end{center}
\end{figure}
\clearpage
\begin{figure}[t]
\begin{center}
FIGURE 2
\end{center}
\label{Fig2}
\begin{center}
\includegraphics[clip,width=0.9\textwidth]{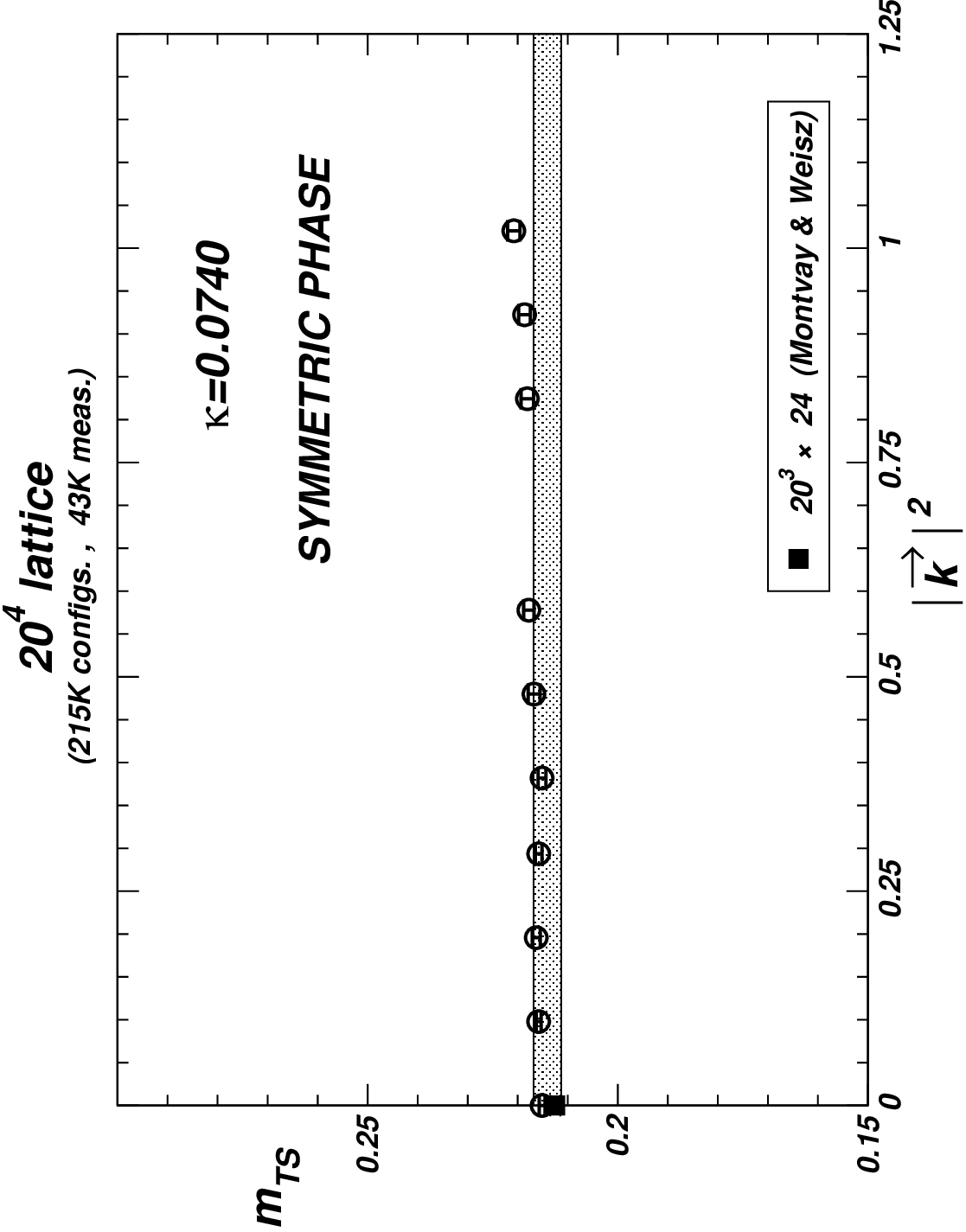}
\end{center}
\end{figure}
\clearpage
\begin{figure}[t]
\begin{center}
FIGURE 3
\end{center}
\label{Fig3}
\begin{center}
\includegraphics[clip,width=0.9\textwidth]{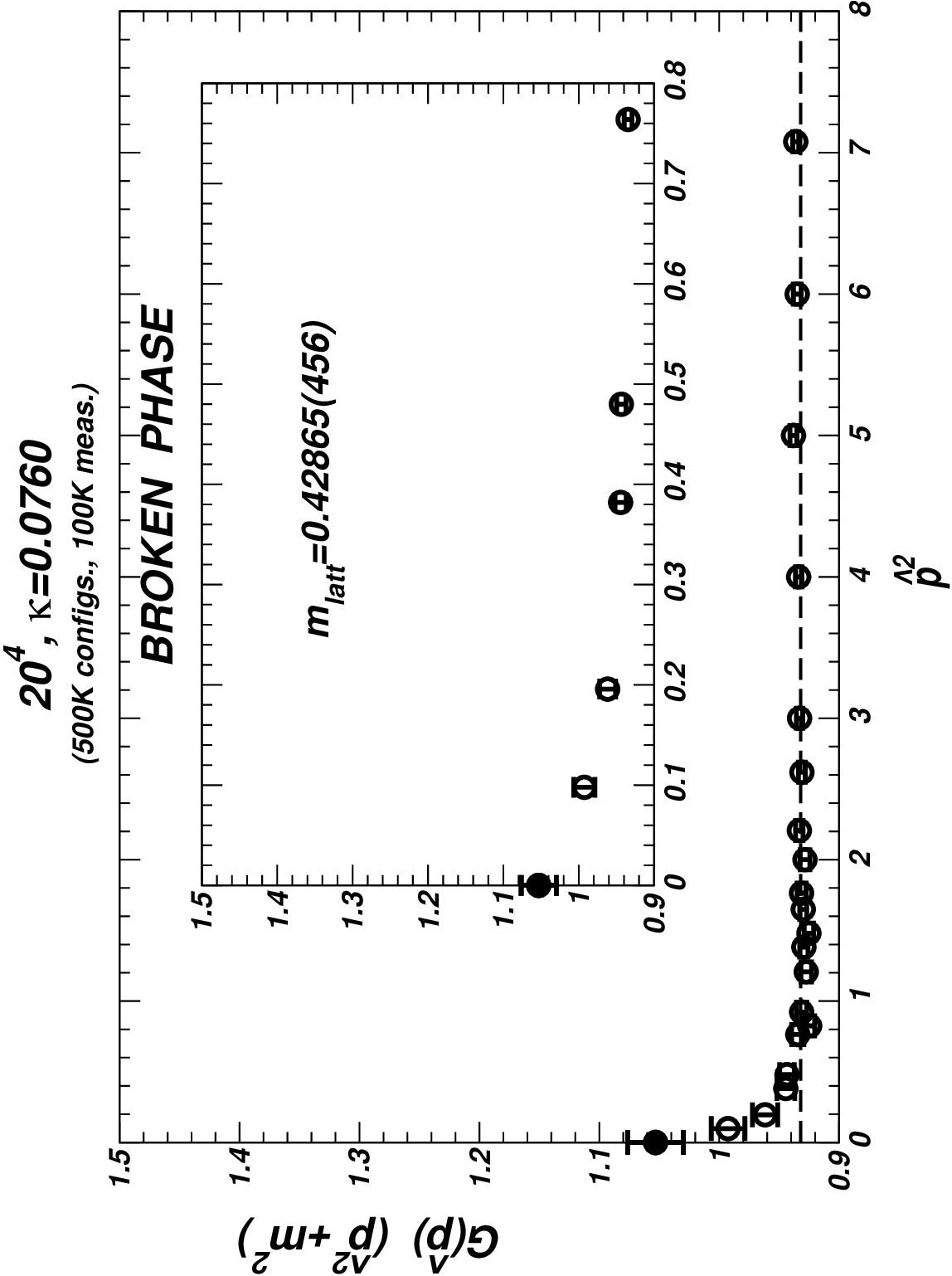}
\end{center}
\end{figure}
\clearpage
\begin{figure}[t]
\begin{center}
FIGURE 4
\end{center}
\label{Fig4}
\begin{center}
\includegraphics[clip,width=0.8\textwidth]{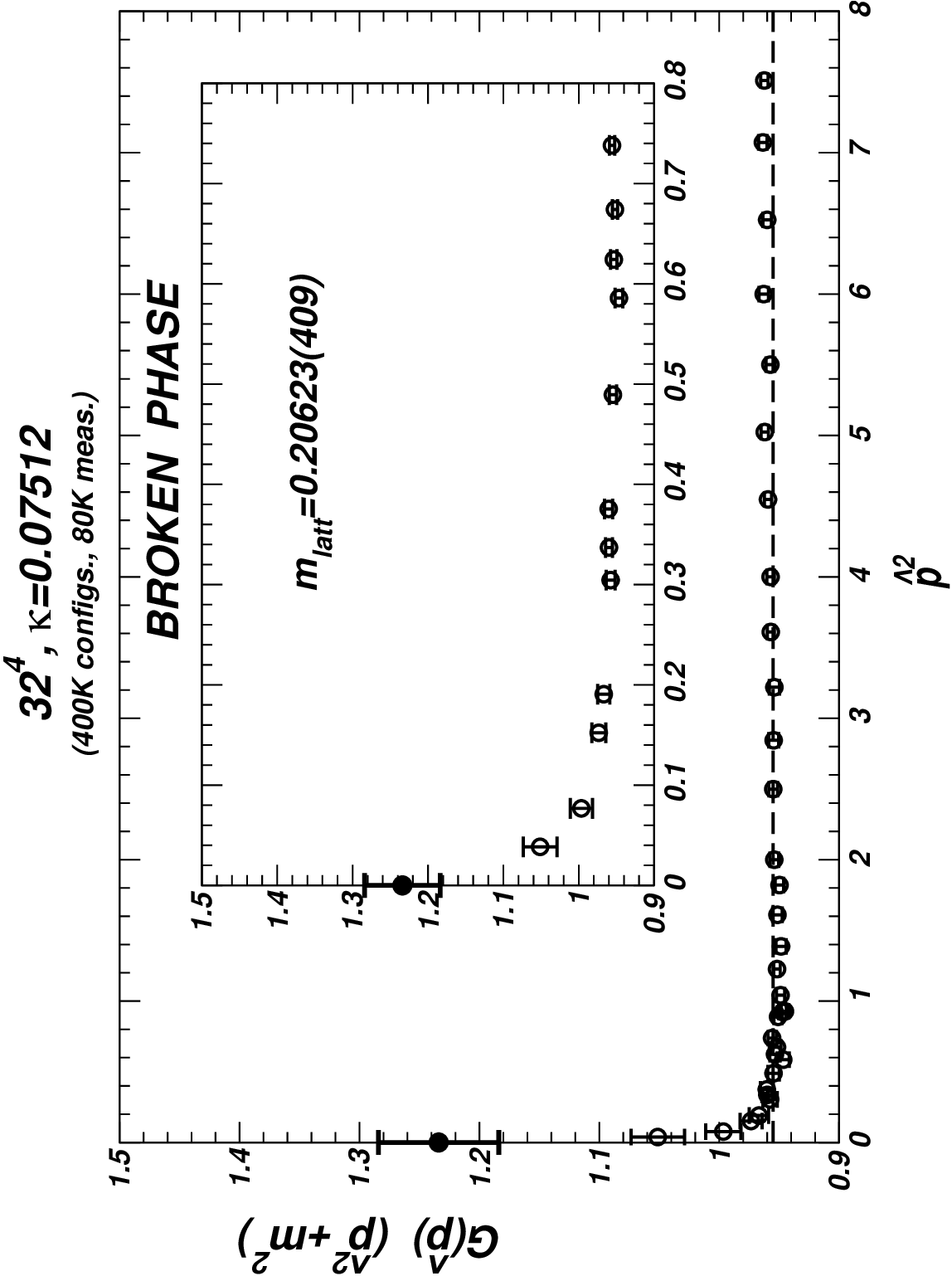}
\end{center}
\end{figure}
\clearpage
\begin{figure}[t]
\begin{center}
FIGURE 5
\end{center}
\label{Fig5}
\begin{center}
\includegraphics[clip,width=0.9\textwidth]{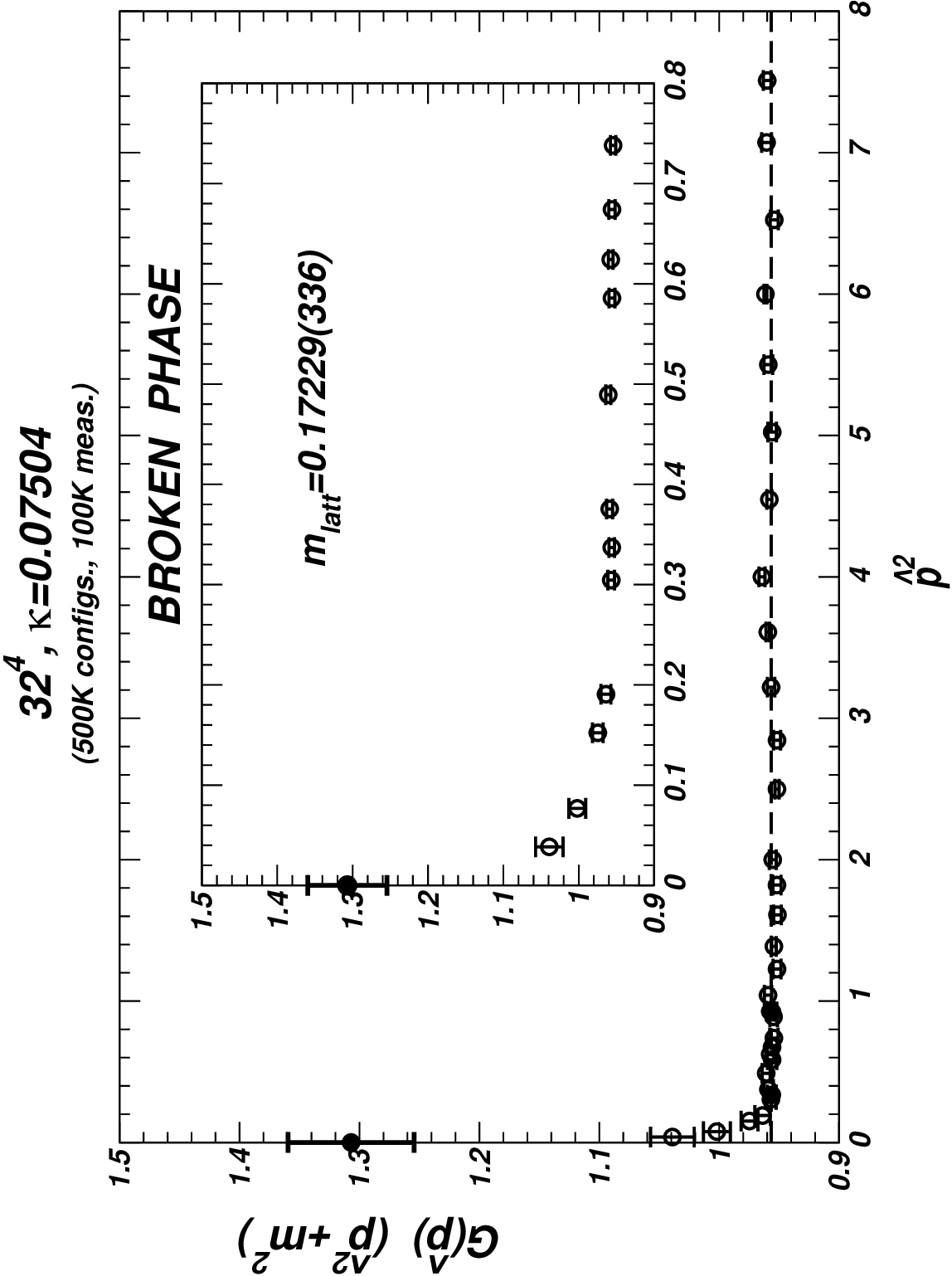}
\end{center}
\end{figure}
\clearpage
\begin{figure}[t]
\begin{center}
FIGURE 6
\end{center}
\label{Fig6}
\begin{center}
\includegraphics[clip,width=0.9\textwidth]{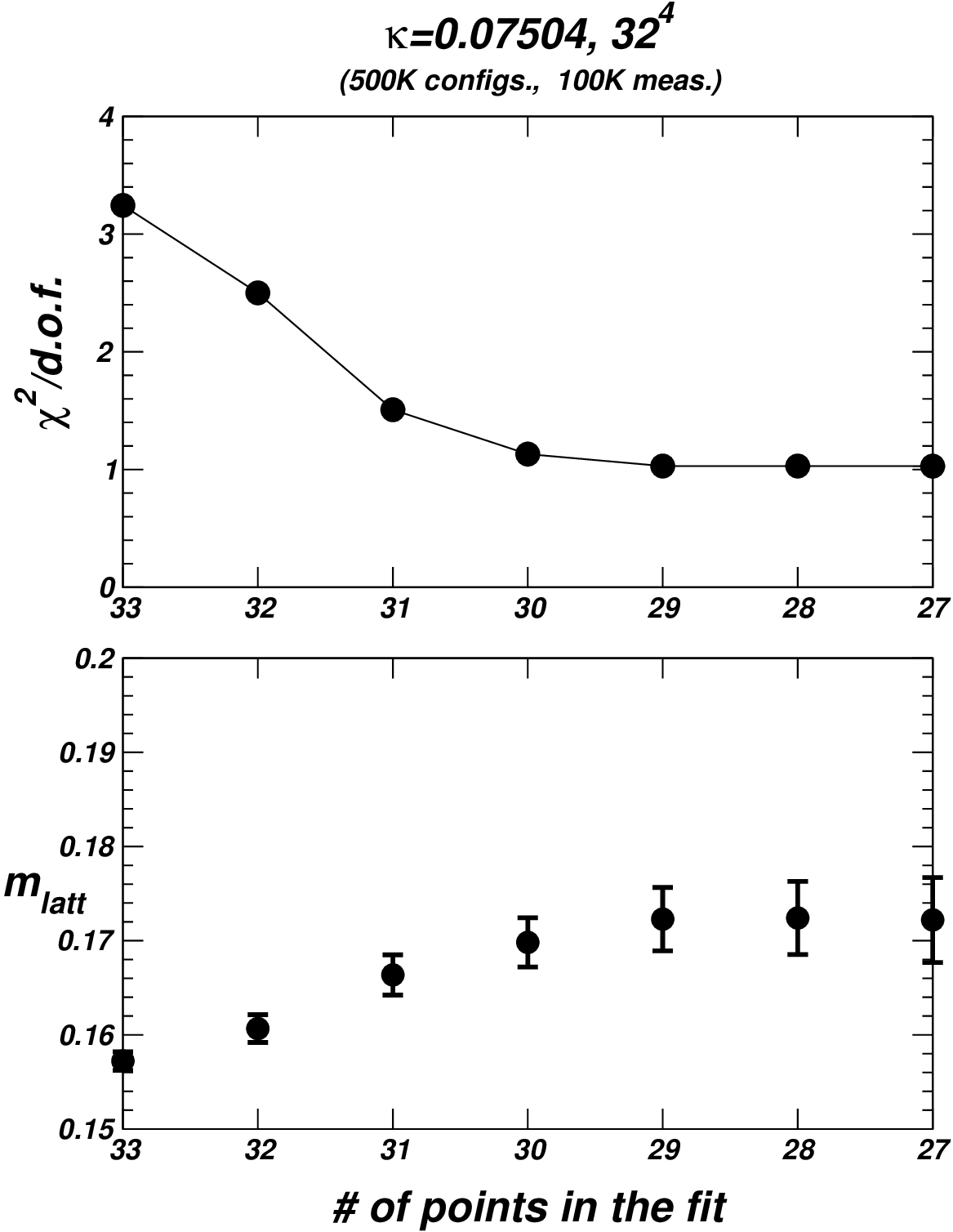}
\end{center}
\end{figure}
\clearpage
\begin{figure}[t]
\begin{center}
FIGURE 7
\end{center}
\label{Fig7}
\begin{center}
\includegraphics[clip,width=0.9\textwidth]{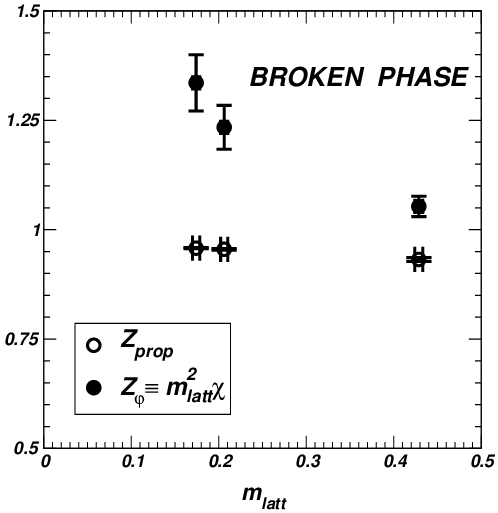}
\end{center}
\end{figure}
\clearpage
\begin{figure}[t]
\begin{center}
FIGURE 8
\end{center}
\label{Fig8}
\begin{center}
\includegraphics[clip,width=0.9\textwidth]{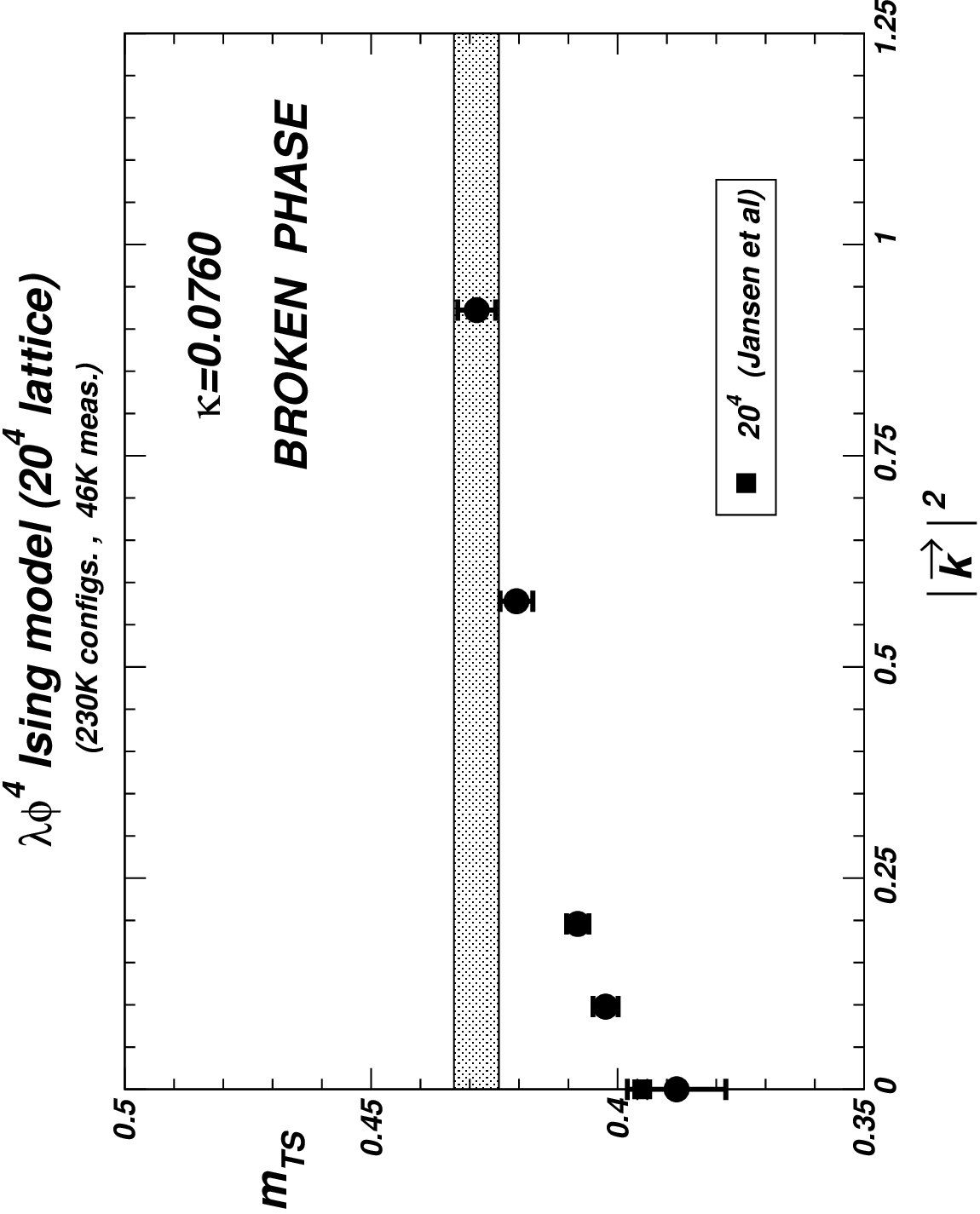}
\end{center}
\end{figure}
\clearpage
\begin{figure}[t]
\begin{center}
FIGURE 9
\end{center}
\label{Fig9}
\begin{center}
\includegraphics[clip,width=0.9\textwidth]{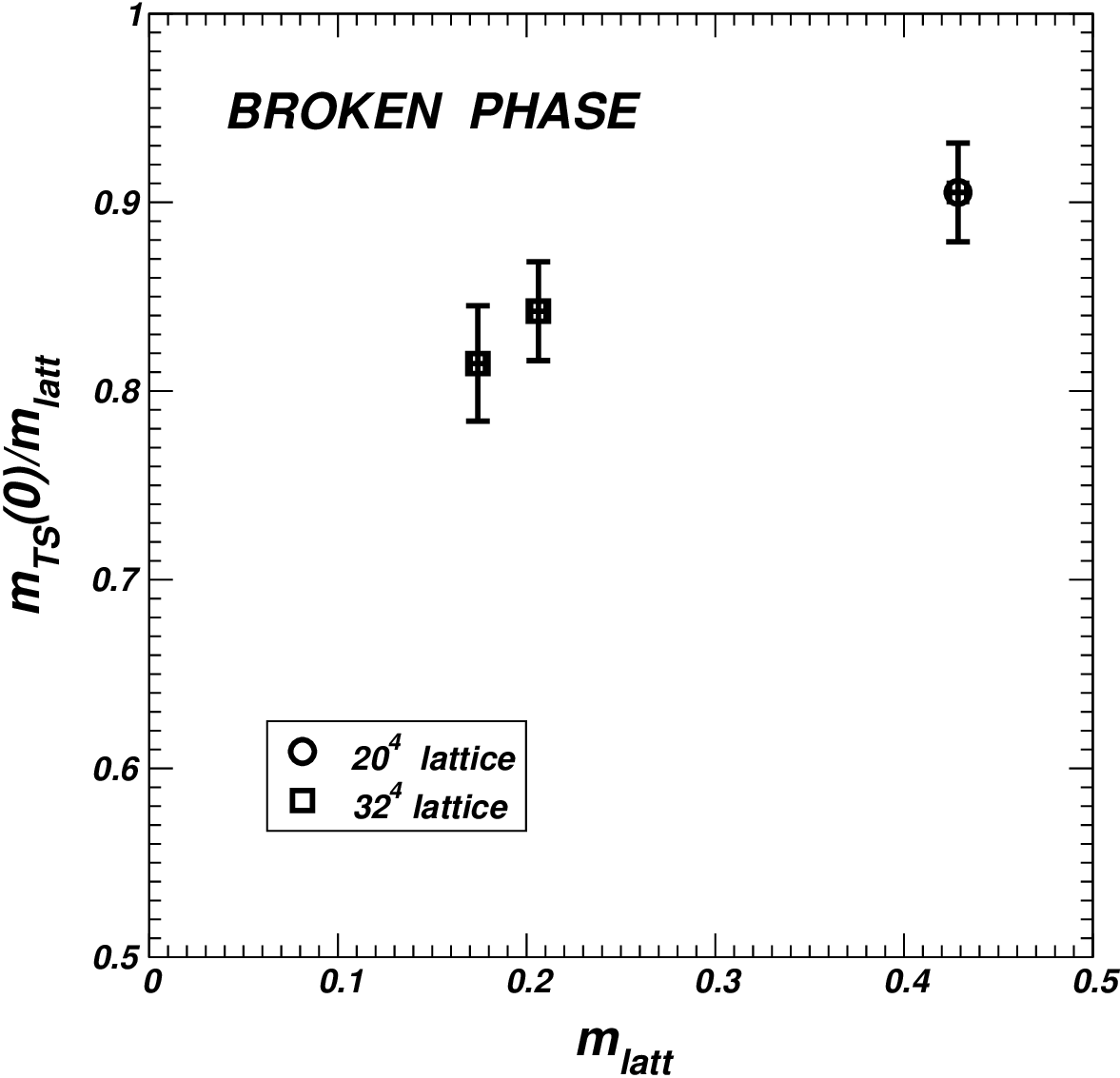}
\end{center}
\end{figure}
\clearpage

\end{document}